\documentclass{aa}  

\usepackage{graphicx}
\usepackage{multirow}
\usepackage{txfonts}
\newcommand{\mj}{$^{-1}$} 
 
\newcommand{\mt}{$^{-3}$} 

\begin{document}

   \title{Properties, age, and origin of a huge meteor cluster observed over Scandinavia on 30 October 2022}

   \author{P. Koten \inst{1}
                  \and
          D. \v{C}apek \inst{1}
          \and 
                  S. Midtskogen \inst{2}
                  \and
                  L. Shrben\'{y} \inst{1}
                  \and
                  P. Spurn\'{y} \inst{1}
                  \and
                  M. Hankey \inst{3}
          }

   \institute{Astronomical Institute, CAS, Fri\v{c}ova 298, 25165 Ond\v{r}ejov, Czech Republic\\
              \email{pavel.koten@asu.cas.cz}
              \and
              Norwegian Meteor Network\\
              \and
              AllSky7 Global Network 
              }

   \date{Received dd-mm-yyyy; accepted dd-mm-yyyy}

 
  \abstract
   {A meteor outburst consisting of at least 22 meteors above the Baltic sea and southern Scandinavia that occurred on 30 October 2022 was recorded using multiple cameras. A bright fireball was followed by fainter meteors over a 10 second period. All the meteors were travelling on parallel trajectories.}
   {The goal of this study is to determine the atmospheric trajectories and photometric masses of meteors and to use these data to determine the specifics of the progenitor meteoroid break-up and cluster formation.}
   {Double and triple-station observations using video cameras were used for the calculation of the atmospheric trajectories and photometric masses of the meteors. Their relative positions and mass distribution were then used to determine the time and cause of the meteoroid fragmentation.}
   {The relative position of the cluster particles in the atmosphere and the distribution of their masses best correspond to the separation of the smaller fragments from the mass-dominant fragment 10.6$\pm$1.7~days before the collision with Earth, assuming a meteoroid bulk density of 1000 kg.m\mt. The ejection velocities are in the range 0.16-0.61~m.s\mj. The directions of the ejection velocities are bounded by a cone with an apex angle of $43^\circ$. The axis of this cone has ecliptic coordinates of $l=154^\circ$ and $b=26^\circ$ and is $66^\circ$ away from the direction to the Sun. Thermal stresses appear to be the most likely cause of such meteor cluster formation.}
   {}


   \titlerunning{Meteor cluster over Scandinavia on 30 October, 2022}
   \authorrunning{P. Koten et al.}

   \maketitle
%

\section{Introduction}

Meteor clusters are typically groups of meteors moving in the same direction and at the same speed, appearing over a relatively short period of time on the order of seconds. Because of their distance from each other, ranging from kilometres to hundreds of kilometres, it is impossible for them to have formed after entering the Earth's atmosphere. The process of their formation is therefore different from the fragmentation that occurs when a meteoroid passes through the Earth's atmosphere. Clusters are therefore considered to be evidence for the fragmentation of meteoroids in heliocentric space. Depending on the separation of the meteoroids, the fragmentation takes place hours or days before the encounter with Earth \citep{Capek2022}.

There are only a few cases of instrumentally recorded meteor clusters. Sometimes it is difficult to distinguish whether the observed phenomenon is a group of physically related meteors or a random fluctuation in the meteoroid flux, especially in the case of very active meteor showers. As the number of meteors observed in a short time interval increases, the likelihood of the presence of a cluster increases. The search for meteor pairs and groups among the Geminid meteors found no evidence for their existence within this meteor shower \citep{Koten2021}. If the cluster is observed during activity of a weak meteor shower or if the meteors are sporadic, the probability of a real connection between the meteoroids increases anew.

Among the already reported cases, there are three meteor clusters observed during the period of Leonid meteor storms at the turn of the century \citep{Kinoshita1999, Watanabe2002, Watanabe2003}. In 2016 a cluster of ten meteors from September $\epsilon$-Perseid shower was observed over the Czech Republic \citep{Koten2017}. A detailed analysis of this cluster by \citet{Capek2022} revealed that thermal stress of fragile cometary material was the most likely process leading to the formation of the cluster. The age of the cluster was found to be about 2.3 days. Since  this cluster will be used more often for comparison, we refer to it as the SPE2016 cluster. Recently, a strong cluster of 38 meteoroids was recorded during a $\tau$-Herculid 2022 meteor outburst \citep{Vaubaillon2023}. In this case, the cluster was observed from only one site and the trajectories of the meteors were not reconstructed.

In this paper we describe another case of multi-station video observation of a meteor cluster. It was recorded by the Norwegian/Allsky7 Bolide networks on 30 October 2022. The first and brightest meteor was observed at 03:54~UT and the last approximately 10~s later. In total, 22 meteors bright enough for analysis were observed\footnote{\url{http://norskmeteornettverk.no/meteor/cluster_20221030/}}. Two others at the noise level turned out to be unmeasurable.

Section \ref{instrument} of this paper describes the instrumentation, data acquisition, and processing procedures. Section \ref{results} summarises the results of the cluster analysis. First, the basic characteristics of the cluster, the atmospheric trajectories, and the heliocentric orbits are described. These results are then used for the detailed analysis of the cluster age and ejection velocities of its members. Finally, the implications in the data for the origin of the meteor cluster are discussed in Section \ref{discussion}.

\section{Instrumentation and data processing}
\label{instrument}

Table~\ref{tab_video_stations} lists the geographical coordinates and number of cameras of all video stations. All 22 meteors observed were captured by the cameras at Gaustatoppen. The cameras in Larvik recorded 17 meteors, while one was recorded by a camera in Oslo. Out of all the meteors, 1 was recorded from three sites, 16 from two sites, and the remaining 5  just from  Gaustatoppen. Moreover, the video camera located on \v{C}erven\'{a} Hora in the Czech Republic also detected the brightest meteor. 

Gaustatoppen station is equipped with AllSky7 cameras \citep{Hankey2020}. Each camera is comprised of five video cameras with an approximate 85-degree horizontal field of view, featuring a 10-degree overlap on either side of the field of view. The two remaining cameras are located in the centre, each having an approximate elevation of 68 degrees. Overall, the system is designed to cover the entire sky from the horizon to the zenith, devoid of any gaps. Currently, the cameras are equipped with the IMX 291 Sony Starvis CMOS chips, providing a resolution of 1920 x 1080 pixels with a 4 mm focal length lens. The cameras were operating at a rate of 25 frames per second. Notably, the Gaustatoppen station camera has been upgraded from the typical AllSky7 cameras and  it is now reinforced with a custom metal dome and additional heating to withstand the harsh weather conditions at the mountain summit.

Larvik station uses four Vivotek IP9171 cameras, which are slightly less sensitive compared to the Allsky7 cameras but have a wider field of view of around 114 x 84 degrees, recording in 2048 x 1536 px resolution. These cameras operated at 15 frames per second. The Vivotek IP8192 model with field of view of 115 x 84 degrees, recording in 2560 x 1920 px and at 5 fps, is used at Oslo station.

The \v{C}erven\'{a} Hora station is one of the European Fireball Network stations. In addition to an all-sky Digital Fireball Observatory, the station features a 4 Mpx Dahua IP camera with a resolution of 2688 x 1520 pixels and 25 frames per second in H.265 encode mode. The camera has a horizontal field of view of 83~degrees. Its northward orientation enabled it to capture a remote meteor above the Baltic Sea. Further details regarding this instrument can be found in \citet{Shrbeny2022}.

\begin{table*}
\caption{Geographical coordinates of the video stations.}             
\label{tab_video_stations}      
\centering          
\begin{tabular}{l r r r r} 
\hline
Station                   &       Longitude E   		&       Latitude N         	&       Altitude    & No. of cameras        \\
\hline
Gaustatoppen              &       8.65566$^{\circ}$     &       59.85112$^{\circ}$  &       1846 m      &       7               \\
Larvik                    &       10.09819$^{\circ}$    &       59.09086$^{\circ}$  &         50 m      &       4               \\
Oslo                      &       10.64964$^{\circ}$    &       59.97056$^{\circ}$  &        348 m      &       3               \\
\v{C}erven\'{a} Hora      &       17.54196$^{\circ}$    &       49.77726$^{\circ}$  &        749 m      &       1               \\
\hline                  
\end{tabular}
\end{table*}

\begin{figure*}
  \resizebox{\hsize}{!}{\includegraphics{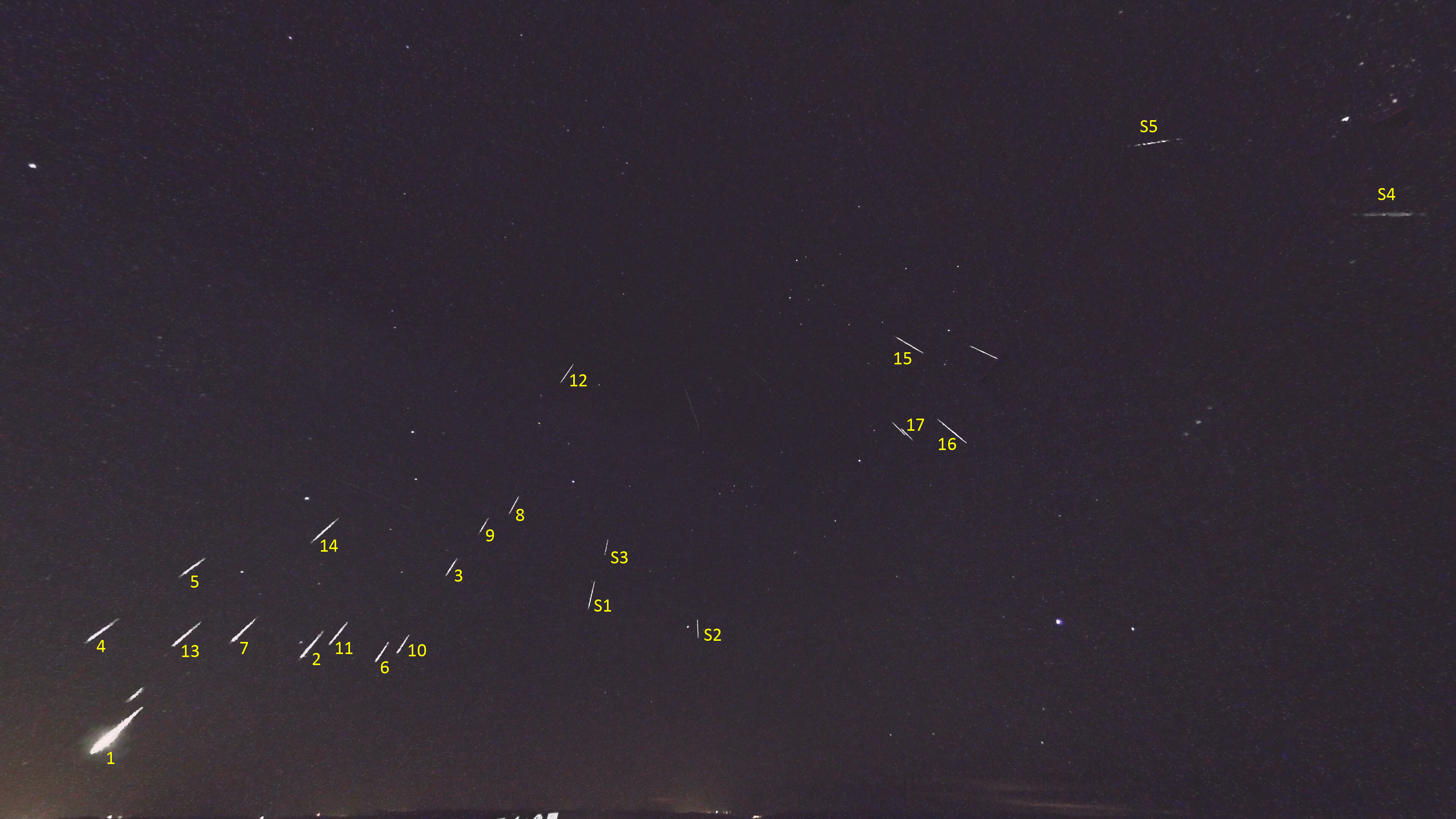}}
  \caption{Recordings from four cameras at Gaustatoppen were combined into a single image in this rectilinear (gnomonic) projection (Credit Steinar Midtskogen). In addition to the numbered multi-station meteors, five single-station meteors are also visible (S1-S5).}
  \label{fig_gaustatoppen_stacked}
\end{figure*}

Preliminary analysis was made by Steinar Midtskogen who carefully studied 17 multi-station meteors frame-by-frame to determine their azimuth and altitude values. Their trajectories were calculated using the {\tt fbspd\_merge.py} tool \citep{Stempels2016}. The results were published on the Norsk Meteornettverk webpage\footnote{http://norskmeteornettverk.no/meteor/cluster\_20221030/} a mere four days after the event occurred, alongside a composite image from four cameras at Gaustatoppen as shown in Figure~\ref{fig_gaustatoppen_stacked}.

Later, all meteor records were measured manually using the FishScan software \citep{Borovicka2022}. The atmospheric trajectories and heliocentric orbits of the double or multi-station meteors were calculated by the Boltrack program \citep{Borovicka2022}. For the videos covering only part of the sky, this program uses the gnomonic conversion formulas from \citet{Borovicka2014b} with a polynomial lens distortion. The atmospheric trajectories of the five single-station meteors were estimated by assuming that their radiant and initial velocities were the same as those of the brightest member of the cluster. These estimations were based on the methods of \citet{Arlt1992} and \citet{Gural1999}. 

Using the atmospheric trajectories, the absolute magnitudes of the meteors were measured at each point. The photometric mass was then calculated using procedure described in \citet{Ceplecha1987}. The luminous efficiency according to \citet{Pecina1983} was used for the integration of the meteor light curves. In the case of a Czech IP camera, photometry was not performed.

The most problematic meteor recorded was also the brightest one due to its occurrence at a considerable distance from the Norwegian cameras which, in turn, resulted in an extremely low convergence angle. In addition, it was captured at a low altitudes above the horizon near the edge of the field of view where only a handful of stars were found around the meteor. This made astrometry even more difficult. Because the Boltrack program allows us to combine star positions measured at different times or even nights for more accurate astrometry, the Norwegian network's archive was searched for other calibration images containing brighter stars in this specific area of the field of view. The images used were taken within one week before or after the cluster was recorded.

Fortunately, the Czech record from a distance of about 1000 km improved the calculation considerably, although the meteor was recorded very low above the horizon. A careful stellar calibration was performed by Luk\'{a}\v{s} Shrben\'{y} using longer exposure time of individual video frames taken on 2 November 2022.

\section{Results}
\label{results}

\subsection{Atmospheric trajectories}
\label{trajectories}

\begin{figure*}
  \resizebox{\hsize}{!}{\includegraphics{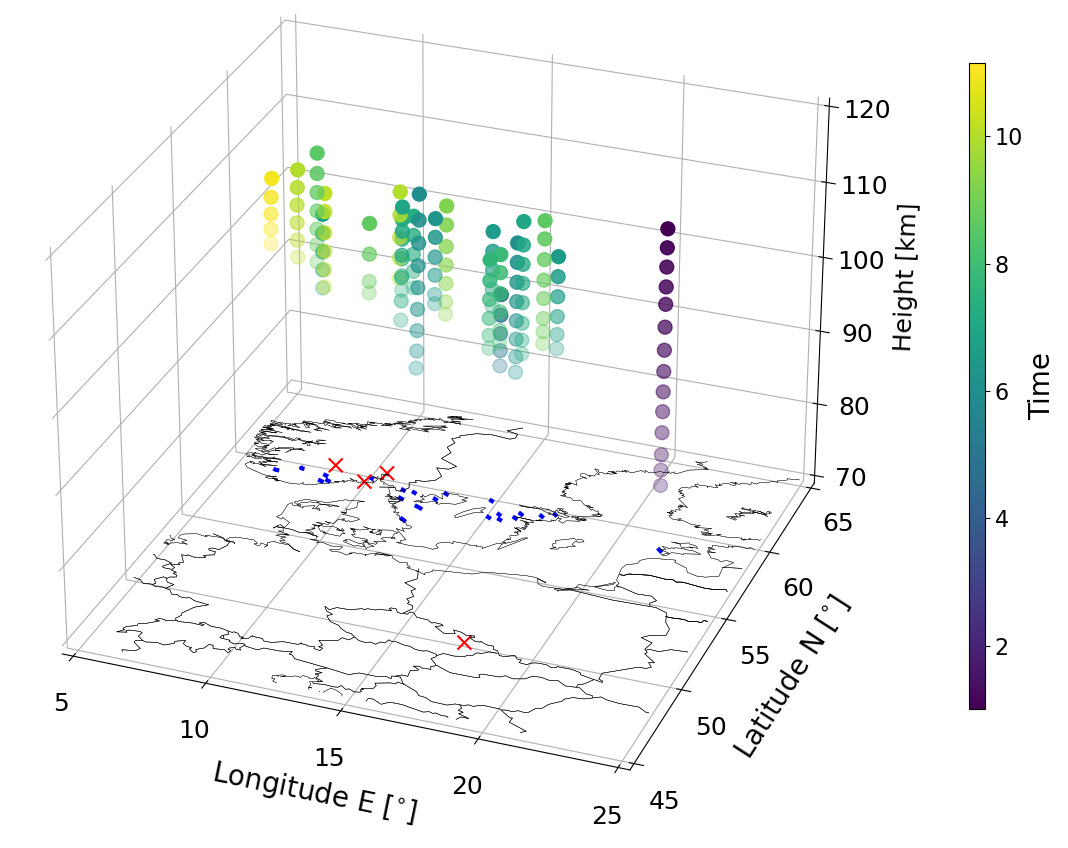}}
  \caption{3D representation of meteor atmospheric trajectories and their projection on a map of Europe. The time from the start of the first meteor expressed in seconds is colour coded. A total of 17 multi-station and 5 single-station meteors are shown. The positions of the stations are plotted as red crosses. The scale on the x and y axes is much smaller than on the z axis, so the projections of meteor trajectories onto the Earth's surface are very short.}
  \label{fig_3D_trajectories}
\end{figure*}

The reconstruction of the atmospheric trajectories of the individual meteors of the cluster shows that the event started at 03:54:04.08~UT over the Estonian island of Saaremaa in the eastern Baltic Sea (Figure~\ref{fig_3D_trajectories}), when the brightest meteor of the cluster was first detected. It was so far from the Norwegian stations -- almost 800 km from Gaustatoppen and about 700 km from Larvik -- that the convergence angle was only 9~degrees. This results in a large uncertainty in the trajectory determination. The accuracy of the trajectory determination was significantly improved by adding the record from the Czech station of the European Fireball Network \v{C}erven\'{a} Hora. Despite the distance of 990~km from the meteor, the nearly perpendicular viewing angle relative to the Norwegian stations made the trajectory determination significantly more reliable. 

Because of the large distance from all stations, the meteor was consistently very low with respect to the horizon. For both Norwegian stations, it had less than 5 degrees in terms of elevation. The observation from the Czech station starts just 1.5~degrees above the horizon and ends almost on the ideal horizon. For this reason a correction for refraction was applied.

The brightest meteor started at an altitude of 115.9~km, reached its maximum absolute brightness of -10.8$^{mag}$ at an altitude of 86.8~km, and terminated its luminous trajectory at an altitude of 74.2~km. The entire trajectory was 43.8~km long and the meteor lasted 0.52~s. Because the video records of the brightest part of the meteor were overexposed, the maximum brightness is rather a lower estimate of the actual value, although some correction on the saturation has been made. The corresponding photometric mass is 123~g. As with the maximum brightness, this is a lower estimate. However, the very similar photometric mass value was found from the video recordings from both Norwegian stations, so it can be considered reliable. Detailed information about the atmospheric trajectory of this meteor and other members of the cluster is given in Table~\ref{tab_atm_traj}.

The second meteor appeared almost 5~seconds later, almost exactly over the Swedish city of Norrk\"{o}ping. Compared to the first meteor, its trajectory was shifted significantly towards the Norwegian stations. The ground distance from the origin of the first meteor was 331~km and the distance from the Gaustatoppen station was about 470~km. This slightly increased the value of the convergence angle to 11~degrees. It was the second most massive meteoroid of the cluster with a photometric mass of 0.38~grams. 

The other meteors followed within another 5.2 seconds with trajectories and ground tracks moved more to the west. Meteors 15, 16, and 17 were recorded at distances of about 150 km south of Gaustatoppen. The last two meteors (single-station meteors) were detected to the southwest of this Norwegian station. The later of these, S4, occurred at 3:54:14.20~UT, and it was the westernmost meteor of the entire cluster. 

The initial points of the first and last meteors were separated by a distance of 880~km when projected on the Earth's surface. The cluster as a whole covered this distance in 10.2~s. This corresponds to a westward velocity of 86.8~km/s. This velocity is orders of magnitude higher than the rotational velocity of the Earth at the corresponding latitude, indicating that the separation of fragments is not due to the rotation of our planet but due to the actual distribution of meteors within the cluster.

The trajectories of all meteors were very steep, with a slope of about 70~degrees. Therefore, their projections on the Earth's surface are very short (as shown in Figure~\ref{fig_3D_trajectories}).

As the cluster moved westward, the convergence angle gradually increased and the calculated trajectories were determined with smaller errors. On the other hand, the fainter meteors were captured in only a few images, especially at the Larvik station, which is equipped with lower frame rate cameras. This, in turn, made the measurements and, subsequently, the trajectory calculations more complicated. Both factors, namely, the small angle of the planes and the low number of measured positions, resulted in relatively higher formal errors for some parameters. Although the positions of the Norwegian stations were unfavourable for the trajectory calculation of the few first meteors, they did enable the observation of the most members of the cluster as it moved from east to west towards the stations. 

\begin{table*}
\caption{Atmospheric trajectories of the multi-station meteors belonging to the 30 October 2022 meteor cluster (J2000.0).}             
\label{tab_atm_traj}
\centering          
\begin{tabular}{l c c c c c c c c c} 
\hline
$met$      &   $time$      &   $H_{BEG}$       &  $\lambda_{BEG}$   &  $\phi_{BEG}$  &  $H_{END}$      &  $l$   &  $t_{dur}$  &  $M_{MAX}$  &  $m_{phot}$   \\
 \#        &   [s]         &   [km]            &  [$^{\circ}$]      &  [$^{\circ}$]  &  [km]           &  [km]  &  [s]        &  [mag]      &  [kg]         \\
\hline
01         &   04.08       &   115.9$\pm$0.3   &  21.89031          &  58.15847      &  74.2$\pm$0.2   &  43.8  &  0.52       &  -10.8      &   1.2e-01     \\   
02         &   08.92       &   114.4$\pm$0.1   &  16.23592          &  58.50805      &  89.2$\pm$0.1   &  26.9  &  0.28       &  -4.6       &   3.8e-04     \\  
03         &   09.32       &   110.5$\pm$0.1   &  13.03281          &  58.78302      &  96.6$\pm$0.1   &  14.9  &  0.20       &  -2.2       &   5.6e-05     \\
04         &   09.52       &   110.0$\pm$0.1   &  17.49133          &  59.26087      &  92.4$\pm$0.1   &  18.6  &  0.20       &  -3.8       &   1.9e-04     \\
05         &   09.68       &   109.3$\pm$0.1   &  14.97878          &  59.40142      &  95.6$\pm$0.1   &  14.6  &  0.18       &  -3.2       &   6.1e-05     \\
06         &   09.96       &   108.8$\pm$0.1   &  15.75913          &  58.20703      &  90.9$\pm$0.1   &  19.7  &  0.19       &  -3.3       &   1.1e-04     \\
07         &   10.00       &   111.9$\pm$0.2   &  16.31756          &  58.86474      &  91.1$\pm$0.2   &  22.2  &  0.28       &  -5.1       &   3.0e-04     \\
08         &   10.20       &   112.2$\pm$0.1   &  11.70162          &  58.99163      &  95.1$\pm$0.1   &  18.3  &  0.22       &  -1.5       &   2.0e-05     \\
09         &   10.32       &   110.0$\pm$0.1   &  12.13192          &  58.96526      &  97.4$\pm$0.1   &  13.5  &  0.16       &  -1.7       &   1.8e-05     \\
10         &   10.60       &   108.7$\pm$0.1   &  15.30498          &  58.22478      &  92.7$\pm$0.1   &  17.4  &  0.20       &  -2.9       &   9.3e-05     \\
11         &   10.84       &   110.1$\pm$0.1   &  15.58398          &  58.55315      &  93.2$\pm$0.1   &  18.0  &  0.20       &  -3.2       &   1.7e-04     \\
12         &   11.60       &   112.0$\pm$0.1   &  10.29596          &  59.39472      &  94.0$\pm$0.1   &  19.3  &  0.28       &  -1.6       &   2.9e-05     \\
13         &   11.64       &   113.0$\pm$0.2   &  17.08626          &  58.97879      &  93.3$\pm$0.2   &  20.9  &  0.28       &  -4.2       &   2.5e-04     \\
14         &   12.20       &   111.6$\pm$0.1   &  13.27000          &  59.28336      &  94.4$\pm$0.1   &  18.3  &  0.24       &  -3.1       &   8.8e-05     \\
15         &   09.72       &   113.2$\pm$0.1   &   8.66083          &  58.99468      &  95.3$\pm$0.1   &  19.4  &  0.16       &  -2.6       &   5.3e-05     \\
16         &   11.64       &   116.2$\pm$0.1   &   8.62990          &  58.55921      &  98.8$\pm$0.1   &  18.8  &  0.27       &  -2.1       &   2.7e-05     \\
17         &   13.16       &   111.2$\pm$0.1   &   8.86014          &  58.66481      &  96.4$\pm$0.1   &  16.1  &  0.20       &  -2.0       &   3.5e-05     \\
\hline
\end{tabular}
\tablefoot{$met$: Meteor number, $time$: Time after 03:54 UT, $H_{BEG}$: Beginning height, $\lambda_{BEG}$ and $\phi_{BEG}$: Longitude and latitude of the beginning point, $H_{END}$: End height, $l$: Length of trajectory, $t_{dur}$: Duration of meteor, $M_{MAX}$: Maximum brightness, $m_{phot}$: Photometric mass.}
\end{table*}

\begin{table}
\caption{Radiants and velocities of the multi-station meteors belonging to the 30 October 2022 meteor cluster (J2000.0).}             
\label{tab_radiants}
\centering          
\begin{tabular}{l c c c c c} 
\hline
$met$           &       $\alpha_{G}$            & $\delta_{G}$      &   $V_{\infty}$         & $V_{G}$               &       $Q$                     \\
 \#                     &       [$^{\circ}$]            &       [$^{\circ}$]    &       [km/s]                  & [km/s]          & [$^{\circ}$]  \\
\hline
01              &       128.7$\pm$0.3           &   42.1$\pm$0.2    &     68.0                   &  67.1$\pm$1.4 &  87.7                 \\   
02              &   128.4$\pm$1.3     &   41.6$\pm$0.5    &     68.9                    &  67.9$\pm$0.3   &               11.2              \\  
03              &   128.3$\pm$1.0     &   42.0$\pm$0.4    &     67.7                    &  66.7$\pm$1.2   &               20.9              \\
04              &   127.2$\pm$2.0     &   42.6$\pm$0.3    &     69.9                    &  69.0$\pm$0.1   &               11.2              \\
05              &   127.7$\pm$2.6     &   42.6$\pm$0.5    &     67.2                    &  66.2$\pm$1.1   &               18.1              \\
06              &   133.2$\pm$1.4     &   39.0$\pm$0.7    &     71.0                    &  70.1$\pm$1.7   &               11.2              \\
07              &   127.2$\pm$0.5     &   42.2$\pm$0.1    &     69.7                    &  68.8$\pm$1.1   &               12.2              \\
08              &   127.2$\pm$0.6     &   42.4$\pm$0.3    &     69.6                    &  68.6$\pm$7.1   &               42.4              \\
09              &   127.8$\pm$1.0     &   41.6$\pm$0.2    &     68.8                    &  67.9$\pm$4.0   &               35.0              \\
10              &   128.7$\pm$2.8     &   40.0$\pm$1.3    &     70.6                    &  69.3$\pm$1.9   &               11.2              \\
11              &   126.3$\pm$0.9     &   42.5$\pm$0.3    &     70.5                    &  69.6$\pm$3.2   &               11.9              \\
12              &   127.8$\pm$0.2     &   42.2$\pm$0.1    &     68.8                    &  67.9$\pm$5.0   &               55.6              \\
13              &   127.2$\pm$4.2     &   41.2$\pm$1.0    &     69.6                    &  68.7$\pm$1.1   &               12.3              \\
14              &   126.8$\pm$0.9     &   42.2$\pm$0.2    &     69.2                    &  68.2$\pm$1.9   &               27.0              \\
15              &   128.3$\pm$0.2     &   42.0$\pm$0.1    &     68.7                    &  67.8$\pm$6.6   &               67.8              \\
16              &   129.2$\pm$0.1     &   41.7$\pm$0.1    &     69.2                    &  68.3$\pm$1.0   &               53.4              \\
17              &   127.5$\pm$0.2     &   41.6$\pm$0.1    &     70.6                    &  69.6$\pm$3.2   &               64.0              \\
\hline
\end{tabular}
\tablefoot{$\alpha_{G}$ and $\delta_{G}$: Coordinates of geocentric radiant, $V_{\infty}$: Entry velocity, $V_{G}$: Geocentric velocity, and $Q$: Convergence angle.}
\end{table}

The limiting meteor magnitude of Allsky7 cameras is around +4.0. For this reason, the value of the $K_{B}$ parameter was corrected by +0.05 \citep{Ceplecha1967}. According to \citet{Cep1988}, the brightest meteor is classified as a type C, which means that the original meteoroid is classified as regular cometary material. With respect to its orbit in the Solar System, it is a subtype of C2, that is long-period comets. The other meteors of the cluster also belong to class C or class B. The mean value $K_{B} = 7.05\pm 0.10$, namely, class C.

\begin{figure}
  \resizebox{\hsize}{!}{\includegraphics{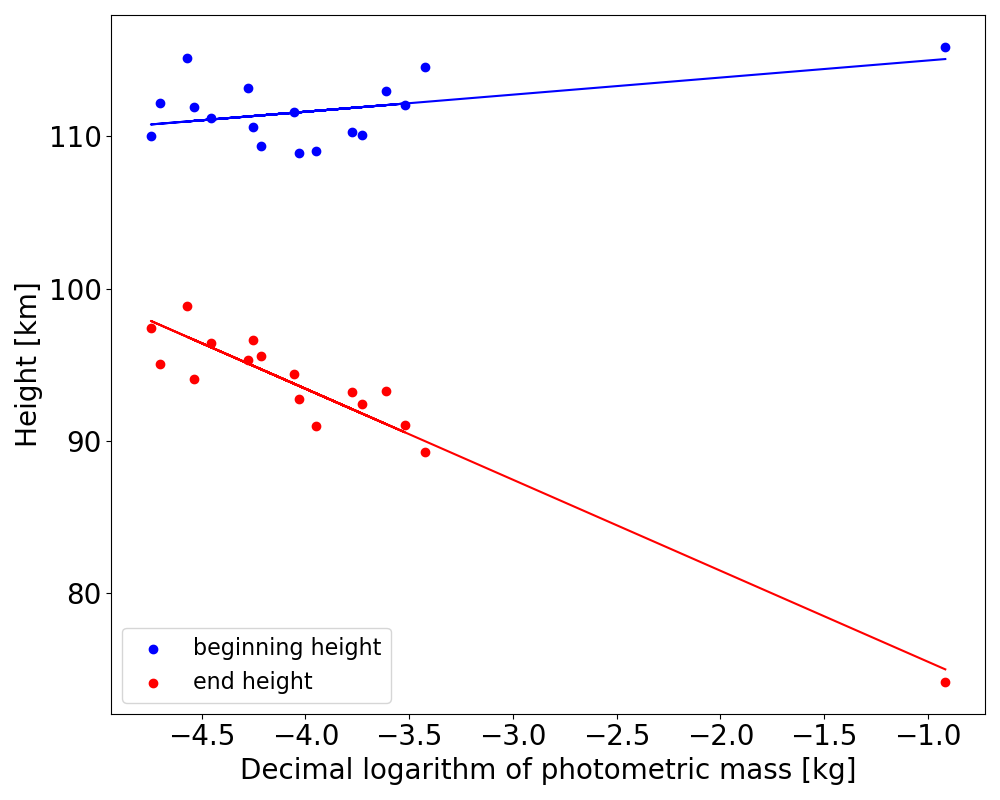}}
  \caption{Beginning (blue) and end (red) heights of 17 multi-station meteors.}
  \label{fig_heights}
\end{figure}

Another clue to the cometary origin of the cluster is the beginning and end heights of the meteors. As Figure~\ref{fig_heights} shows, the trend in the dependence of both heights on photometric mass is similar to that of meteor showers of cometary origin. Such showers are characterised by an increase in the beginning height and a decrease in the end height with increasing photometric mass \citep{Koten2004}.

\subsection{Heliocentric orbits}
\label{orbits}

In addition to the atmospheric trajectories, the heliocentric orbits were calculated for the double-station meteors. Table~\ref{tab_hel_orbit} presents the heliocentric orbit of the brightest meteor of the cluster.

The working list of the IAU Meteor Data Center was searched to determine the possible shower membership of the brightest meteor in the cluster. The closest meteor shower according to Drummond's $D_{d}$-criteria is the October Lyncid shower\footnote{228 OLY code in IAU MDC database}. The value of 0.11 is just above the usually accepted threshold for the shower association \citep{Drummond1981}. Because the radiant differs by almost 14$^{\circ}$, it is likely a sporadic meteor. On the other hand, little data is available about this meteor shower. Its activity is centred on 19 October, 11 days before the cluster was recorded. The orbit of the shower is based on only six meteors \citep{Jenniskens2006}. However, the daily motion of the radiant is unknown. Thus, it is not possible to determine whether the radiant has moved towards the cluster radiant as a result. Moreover, Southworth-Hawkins criterion $D_{SH}$ \citep{Southworth1963} did not provide any close association.

\subsection{Positions and masses of fragments}
\label{PosMass}

The determination of the relative positions and masses of the individual cluster members was performed in a similar way to the method presented in \citet{Capek2022}. A detailed description of this procedure can be found there. In the present text, we also refer to the cluster members as `fragments'.

The photometric masses of the individual fragments were determined from the corresponding light curves using the \citet{Pecina1983} formula for luminous efficiency. The fragment ``01'' has a mass of approximately $120$~g, while the other members have masses several orders of magnitude smaller, in the range $0.39$--$0.016$~g as can be seen in Table~\ref{tabPosMas}. Similar to the case of SPE2016 cluster \citep{Koten2017}, there is a mass-dominant fragment, which can be considered as the rest of the cluster's parent meteoroid, and a group of small fragments that altogether represent only a small fraction of the mass of this parent body. 

\begin{table*}[h]
\caption{Properties and ejection velocities of individual fragments.}
\label{tabPosMas}
\centering
\begin{tabular}{c r@{} l r r r r r | r@{} l} 
\hline 
 met &
 \multicolumn{2}{c}{$m$} &
 \multicolumn{1}{c}{$d$} &
 \multicolumn{1}{c}{$X$} &
 \multicolumn{1}{c}{$Y$} &
 \multicolumn{1}{c}{$Z$} &
 \multicolumn{1}{c}{2$\sigma^{\rm R}$} &
 \multicolumn{2}{c}{ejection velocity} \\
\\
 \#       &
 \multicolumn{2}{c}{[$10^{-3}$g]} &
 \multicolumn{1}{c}{[mm]} &
 \multicolumn{1}{c}{[km]} &
 \multicolumn{1}{c}{[km]} &
 \multicolumn{1}{c}{[km]} &
 \multicolumn{1}{c}{[km]} &
 \multicolumn{2}{c}{[m\,s\mj] } \\
\hline

01 &    123&\raisebox{0.3ex}{$^{+   513}_{-    99}$}g &  61.7&     0.0&   0.0&     0.0&     8.7  &            &                          \\                
02 &    385&\raisebox{0.3ex}{$^{+   175}_{-   120}$} &   9.0&   357.2&  214.9&   103.7&     9.7        &       0.26 & $\pm$  0.05\\ 
03 &     56&\raisebox{0.3ex}{$^{+    11}_{-     9}$} &   4.7&   535.9&  168.2&   109.5&    34.2        &       0.40 & $\pm$  0.08\\ 
04 &    187&\raisebox{0.3ex}{$^{+    51}_{-    40}$} &   7.1&   285.3&  214.4&   193.9&    22.5        &       0.37 & $\pm$  0.07\\ 
05 &     61&\raisebox{0.3ex}{$^{+    18}_{-    14}$} &   4.9&   423.4&  165.8&   182.9&    13.4        &       0.39 & $\pm$  0.08\\ 
06 &    113&\raisebox{0.3ex}{$^{+    25}_{-    20}$} &   6.0&   390.9&  163.7&    49.7&    20.2        &       0.26 & $\pm$  0.05\\ 
07 &    301&\raisebox{0.3ex}{$^{+   422}_{-   176}$} &   8.3&   355.9&  253.4&   160.7&    13.3        &       0.34 & $\pm$  0.06\\ 
08 &     20&\raisebox{0.3ex}{$^{+     4}_{-     3}$} &   3.4&   612.4&  179.8&   139.6&    21.8        &       0.54 & $\pm$  0.10\\ 
09 &     18&\raisebox{0.3ex}{$^{+     4}_{-     3}$} &   3.3&   589.6&  195.5&   142.5&    16.8        &       0.61 & $\pm$  0.11\\ 
10 &     93&\raisebox{0.3ex}{$^{+    21}_{-    17}$} &   5.6&   420.2&  318.1&   108.5&    31.4        &       0.42 & $\pm$  0.08\\ 
11 &    168&\raisebox{0.3ex}{$^{+    36}_{-    30}$} &   6.8&   404.8&  314.3&   146.8&    31.7        &       0.40 & $\pm$  0.07\\ 
12 &     29&\raisebox{0.3ex}{$^{+     7}_{-     6}$} &   3.8&   693.5&  204.6&   199.5&    46.4        &       0.39 & $\pm$  0.08\\ 
13 &    246&\raisebox{0.3ex}{$^{+    74}_{-    57}$} &   7.8&   325.0&  356.3&   218.9&    33.8        &       0.47 & $\pm$  0.09\\ 
14 &     88&\raisebox{0.3ex}{$^{+    49}_{-    31}$} &   5.5&   538.8&  307.8&   227.4&    31.0        &       0.43 & $\pm$  0.08\\ 
15 &     53&\raisebox{0.3ex}{$^{+    28}_{-    18}$} &   4.7&   775.1&  86.6&   106.8&     4.6\       &       0.16 & $\pm$  0.03\\ 
16 &     27&\raisebox{0.3ex}{$^{+     8}_{-     6}$} &   3.7&   796.3&  229.2&   115.6&    18.1        &       0.33 & $\pm$  0.06\\ 
17 &     35&\raisebox{0.3ex}{$^{+    11}_{-     8}$} &   4.1&   794.1&  326.8&   165.4&    17.5        &       0.41 & $\pm$  0.07\\ 
S1 &     86&\raisebox{0.3ex}{$^{+    45}_{-    30}$} &   5.5&   576.6&  167.6&    37.3&    14.5        &       0.20 & $\pm$  0.04\\ 
S2 &     58&\raisebox{0.3ex}{$^{+    29}_{-    19}$} &   4.8&   594.3&  320.9&   -32.2&    22.6        &       0.38 & $\pm$  0.07\\ 
S3 &     56&\raisebox{0.3ex}{$^{+    28}_{-    19}$} &   4.7&   605.6&  486.8&   158.4&    26.0        &       0.58 & $\pm$  0.10\\ 
S4 &     42&\raisebox{0.3ex}{$^{+    25}_{-    16}$} &   4.3&   868.9&  314.3&   196.9&    69.7        &       0.42 & $\pm$  0.09\\ 
S5 &     16&\raisebox{0.3ex}{$^{+     4}_{-     3}$} &   3.1&   854.6&  254.4&   197.9&    63.3        &       0.47 & $\pm$  0.10\\ 

\hline
\end{tabular}
\tablefoot{The photometric pre-atmospheric mass $m$ has two combined standard uncertainty intervals. The diameter, $d,$ is given for a density of $\rho=1000$~kg\,m\mt\ . The coordinates $(X,Y,Z)$ are expressed with respect to the system that is connected with the most massive fragment `01'. The  $x$-axis points in the antisolar direction, and the $z$-axis points to the north pole of the ecliptic. Lastly, $\sigma^{\rm R}$ represents the combined standard uncertainty of positions. The right part of the table shows calculated mean values of velocity with two standard combined uncertainties.}
\end{table*}

\begin{figure*}[t]
    \centering
        \begin{tabular}{l@{\hspace{5mm}}r}
                        \includegraphics[height=7cm]{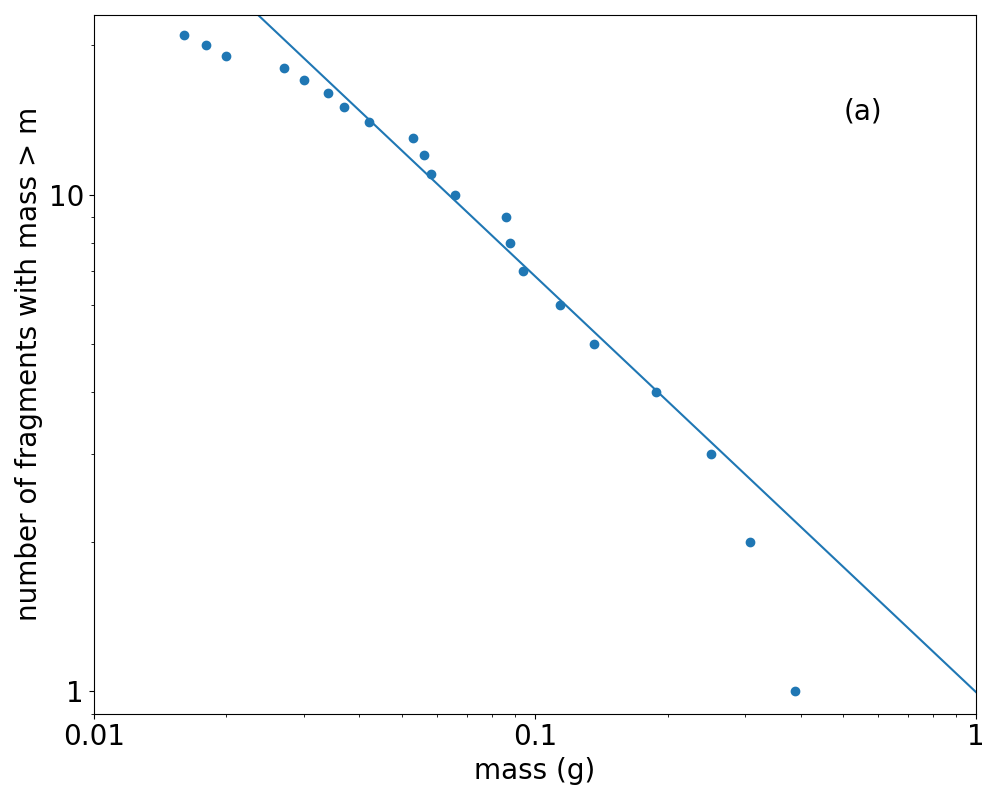}&
                        \includegraphics[height=7cm]{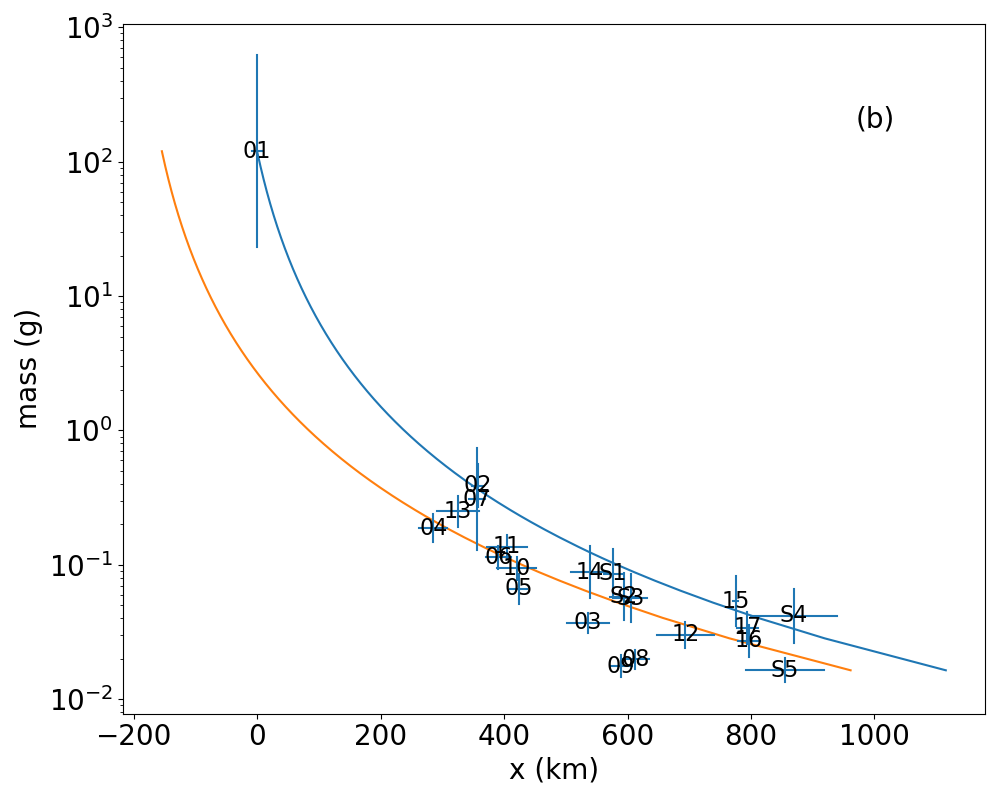}\\
                \end{tabular}
    \caption{Mass distribution and shift of fragments.
            (a) Cumulative mass distribution of fragments. The line corresponds to $N(>m)\propto m^{-0.84}$, i.e. $D=2.52$, and it is the fit for fragment masses in the range of $0.03$--$0.3$~g.
            (b) Shift of fragments from the parent meteoroid in the antisolar direction as a function of mass. The error bars correspond to uncertainty intervals of $2\sigma$. The blue curve corresponds to the theoretical displacement of fragments due to solar radiation pressure for a cluster age of $10.6$~days if the ejection velocities were zero. The deviations of the fragment positions from this curve are caused by the projection of the non-zero ejection velocity into the solar/antisolar direction. If all ejection velocities were $0.39$~m\,s\mj\ in the direction of the ejection cone axis (i.e., $66^\circ$  from the direction to the Sun), the theoretical dependence of the displacement on mass would follow the orange curve.}
    \label{fm}
\end{figure*}

As in \citet{Capek2022}, we consider the possibility that the cluster also contained smaller fragments. However, the corresponding meteors were so faint that they could not be detected. We assume that the number of recorded fragments with masses above $m_{lim}$ = 0.03~g is complete. The total mass below this limit must be estimated using the procedure described in \citet{Capek2022}. The number of fragments larger than a given size $s$ can be expressed using a power function with an exponent $D$ as $N(>s)\propto s^{-D}$. For the cumulative number of fragments with mass greater than $m$, we can write $N(>m)\propto m^{-D/3}$. For masses in the interval $0.03$--$0.3$~g, this distribution can be approximated by the function $N(>m)\propto m^{-0.84}$, namely, with a coefficient $D = 2.52$ (see Figure~\ref{fm}a). Using fragments with smaller masses\footnote{This means all except for fragment ``01''.} from Table~\ref{tabPosMas} and Eq.(3) from \citet{Capek2022}, we can estimate that the total mass of ejected fragments is $5.98$~g, which is $5\%$ of the mass of the parent meteoroid. Total mass of observed fragments $2.115$~g represents about $35\%$ of total ejected mass. 

We determined the positions of the individual fragments using the method described in \citet{Capek2022} and expressed them in the coordinate system connected with the most massive body `01'. The $x$-axis of this system points to the antisolar direction and the $z$-axis points toward the north pole of the ecliptic. The specific positions and their errors are listed in Table~\ref{tabPosMas}. The cluster is very large; smaller fragments occupy a volume of $585 \times 400 \times 260$~km, which is somewhat elongated in the $x$-axis direction. This region is shifted by $285$~km in the antisolar direction with respect to the main fragment. Taking into account the position of the individual cluster members within the field of view (and also because AllSky7 is an all-sky system), we assume that all observable fragments of the cluster have been recorded and that none of its members are outside the field of view or below the horizon. 

\subsection{Cluster age and ejection velocities}
\label{AgeVelocities}

Similar to the case of the SPE2016 cluster, we see that the smaller fragments are shifted away from the mass-dominant fragment, which we consider to be the parent meteoroid, in the antisolar direction. This shift is larger for the less massive fragments and vice versa (see Figure~\ref{fm}b). This may indicate longer action of solar radiation pressure. We assume that the resulting cluster arrangement is the result of two factors: (i) the action of solar radiation pressure and (ii) the cluster formation process itself, which is manifested by specific ejection velocity vectors of individual fragments. We used the method of \citet{Capek2022} to determine both the ejection velocities and the cluster age. This method requires knowledge of the positions, masses, and corresponding uncertainties of fragments at a specific time of observation and it also requires us to establish the time at which the total kinetic energy of freshly ejected fragments reaches its minimum. A more detailed discussion is given in Section~\ref{formation}.

\begin{figure*}[t]
    \centering
        \begin{tabular}{l@{\hspace{5mm}}r}
                        \includegraphics[height=7cm]{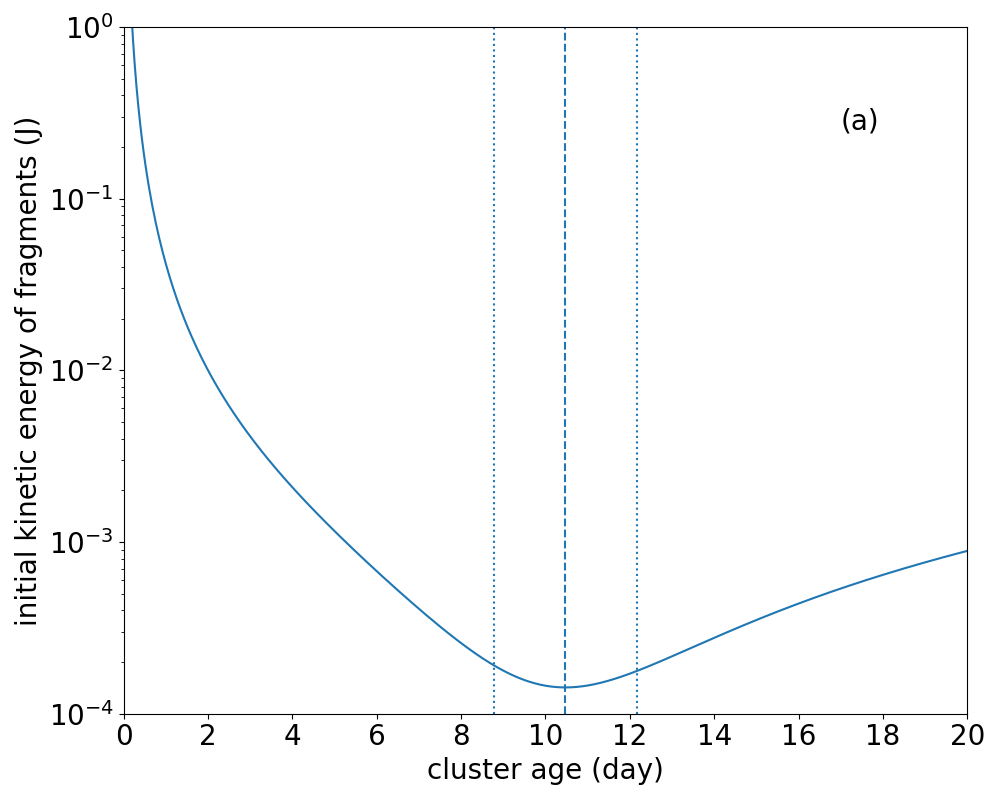}&
                        \includegraphics[height=7cm]{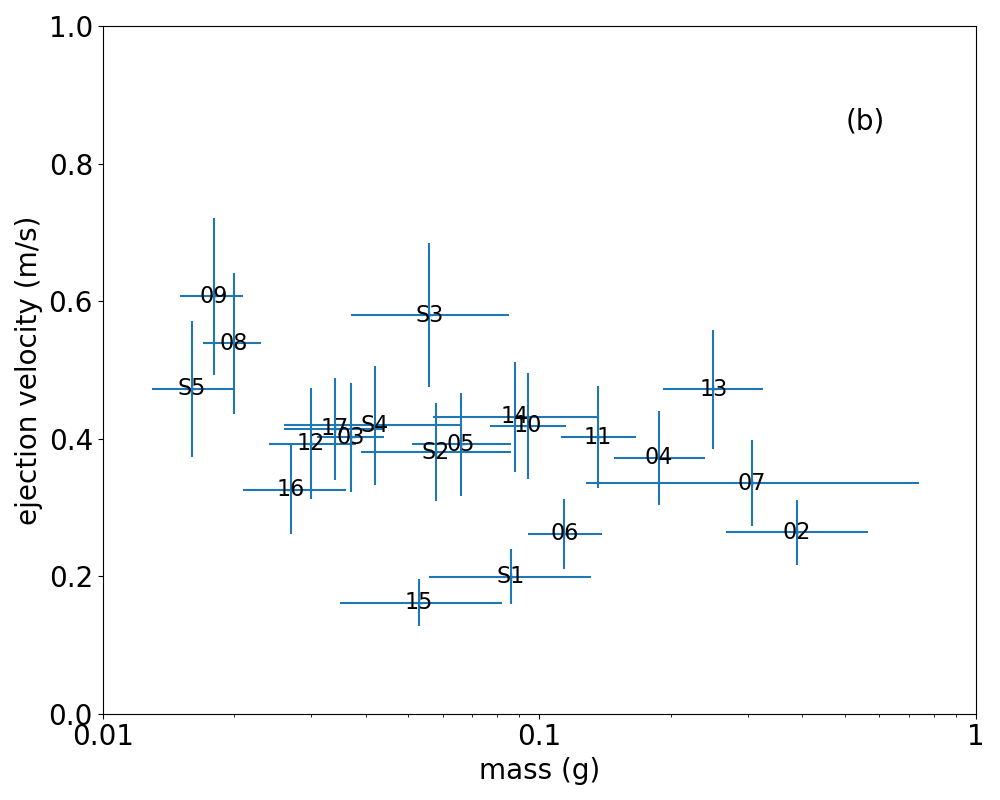}\\
                \end{tabular}
    \caption{Initial kinetic energy and ejection velocities of fragments.
            (a) Total initial kinetic energy of fragments as a function of the age of the cluster. The dashed line corresponds to the cluster age and dotted lines denote two standard combined uncertainties per side.
            (b) Ejection velocity of fragments as a function of mass. The error bars indicate two standard combined uncertainties per side from each point.}
    \label{Ekv}
\end{figure*}

With the assumed density of $1000$~kg\,m\mt\ and the value of the shape parameter $A=1.6$, the age is $10.56$~days (see Figure~\ref{Ekv}a) with a $2\sigma$ uncertainty interval of $0.49$\,days due to photometric mass and position uncertainties. Individual meteors can be classified according to the $K_{\rm B}$ parameter mostly into classes B and C \citep{Cep1988}. Thus, we assume that the density of meteoroid material may be in the range $700$--$1200$~kg\,m\mt. The corresponding ages are $9.37$~days and $11.22$~days. The assumed value of the luminous efficiency is also uncertain and we can consider that it could be half or twice as high. The photometric masses will also change by the same factor and the cluster age will then be $9.40$~days and $11.85$~days. Taking all of these values into account, the cluster age is $10.6 \pm 1.7$~days (a two-standard combined uncertainty). If we take into account the velocity of the main fragment before entering the Earth's atmosphere, we find that the meteoroid disintegration occurred at a distance of 62$\pm$10 million kilometers from the intersection of the orbits of the both bodies.

The initial ejection velocities of fragments are in the range  $0.16$--$0.61$~m\,s\mj (see Table~\ref{tabPosMas} and Figure~\ref{Ekv}b), with a mean value of $0.39$~m\,s\mj. Similar velocities, below $1$~m\,s\mj, were determined for SPE2016 cluster \citep{Capek2022}. 

The directions of the ejection velocities are inside a cone with an apex angle of $43^\circ \pm 11^\circ$. The axis of this cone has ecliptical coordinates $l=154^\circ$ and $b=26^\circ$ and is $66^\circ$ away from the direction to the Sun. The mean direction from which small meteoroids impact the parent meteoroid \citep[calculated using NASA's Meteoroid Engineering Model 3,][]{Moorhead2020} makes an angle of $147^\circ$ with this axis.

\section{Discussion}
\label{discussion}

The reported cluster is one of the few well-documented cases of meteor clusters. The atmospheric trajectories of most meteors have been determined with good precision. It is the second case for which a detailed analysis has been carried out to determine its characteristics and formation circumstances. If we compare it with the SPE2016 cluster, we see that it is much bigger, namely: its dimensions are about $10\times$ larger with twice the number of observed fragments.

\subsection{Comparison with SPE2016 cluster}
\label{SPE2016comparison}

Notwithstanding the the difference in cluster size and fragment count, both clusters are very similar. Both contain a mass-dominant fragment that can be considered as a parent meteoroid from which a small fraction of the mass represented by smaller fragments has separated. These smaller fragments are in both cases significantly shifted in the antisolar direction.

The studied cluster has roughly twice as massive mass-dominant fragment compared to SPE2016. The other observed fragments are about ten times more massive compared to SPE2016; while smaller fragments have not been detected due to the different sensitivity of the MAIA and AllSky7 systems. It is very interesting that the fragment mass distribution can be described by an almost identical power law in both cases: $N(>m) = m^{-0.85}$ for SPE2016 cluster and $N(>m) = m^{-0.84}$ for the current cluster.

Taking into account the much more spatially distant fragments of this cluster, it is not surprising that its age is much higher, $10.6 \pm 1.7$~days, in comparison with $2.28 \pm 0.44$~days for the SPE2016 cluster. The determination procedure of ages and ejection velocities by \citet{Capek2022} is limited by the condition that this age is much smaller than the orbital period. This condition is satisfied, but the problem is that it should be formulated differently for orbits with high eccentricity: during the time interval from the formation of the cluster to its collision with the Earth, there is only a negligible change in the heliocentric position of the cluster. Given this condition, the simple theory of \citet{Capek2022} is at the boundary of usability, noting that the trajectories of the individual fragments would have to be numerically integrated. In our case, however, the corresponding errors are within the uncertainty interval.

The ejection velocity range of $0.16$ -- $0.61$~m\,s\mj\ , with a mean value of $0.39$~m\,s\mj\ , is very similar to those of the SPE2016 cluster ($0.13$ -- $0.77$~m\,s\mj\ with mean value of $0.35$~m\,s\mj ). The only difference is that they are ejected at a much narrower angle (ejection cone of apex angle $43^\circ$ vs. $101^\circ$) and in a direction that does not correspond to the flux of small meteoroids to the parent meteoroid ($147^\circ$ away vs. $34^\circ$ away).

\subsection{Formation process}
\label{formation}

Here, we focus on the question of how the cluster was formed. First, we consider the possibility that the cluster was formed by the impact of a small meteoroid on the parent meteoroid without catastrophic disruption, while only ejecting a small part of the mass. As in the case of SPE2016, it is very difficult to explain the low ejection velocities of the observed fragments. For a more detailed discussion, we refer to \citet{Capek2022}.

Higher ejection velocities can be achieved only if the age of the cluster does not correspond to the time for which the total kinetic energy of the ejected fragments reaches a minimum value. We assume that the mean ejection velocity of $2$~m\,s\mj\ is compatible with the impact origin of the cluster. This value can be reached at ages either higher than $30$~days or, on the other hand, lower than $3.3$~days. Ejection velocities of fragments for a 30-day old cluster would be in the range $1.0$ -- $3.2$~m\,s\mj . Their directions would be inside a cone with an apex angle of only $3.7^\circ$, with an axis $3.5^\circ$ degrees from the Sun. Moreover, there would be a significant dependence of the ejection velocity on the fragment mass $v \propto m^{-0.37}$ with a minimal scatter. Such a configuration is highly unlikely and can be reliably ruled out. 

Ejection velocities of fragments for a 3.3-day old cluster would be in the range of $1.3$ -- $3.1$~m\,s\mj . Their directions would be inside a cone with an apex angle of $23^\circ$, with an axis that is $149^\circ$ degrees from the Sun. It is known from laboratory experiments that impactors produce ejection cones with an apex angle of about $90^\circ$ and with the axis usually perpendicular to the target surface. We would therefore expect the apex angle to be four times larger and the deviation from the mean radiant of the incoming impactors to be within $90^\circ$. The derived distribution of ejection velocity directions is not compatible with this expectation. An impact origin with low age is therefore also unlikely.

We go on to consider the possibility that the cluster was formed by the separation of a part of the mass due to fast rotation. To study this possibility, we can imagine a simplified situation of a fragment of a simple shape (e.g. a cube) on the equator of a rotating parent meteoroid. We can then compare the stress at its base with the expected strength of the material. This stress increases with the mass of the fragment, so we can assume that the mass of all fragments (both observed and those that produced too faint meteors) is concentrated in a single fragment of mass $2.115$~g. With a density of $1000$~kg\,m\mt, the diameter of the parent meteoroid is about $61$~mm (Table \ref{tabPosMas}). We assume that the circumferential velocity corresponds to the highest ejection velocity, which is $0.61$~m\,s\mj.\   The rotational frequency is then $3$~Hz. According to Eq.~(15) in \citet{Capek2022}, the corresponding mechanical stress due to centrifugal force is then $\sim 160$~Pa. This value is about $360$ times lower than the mechanical strength of the material of the parent meteoroid, if we estimate it from the value of the highest dynamic pressure\footnote{We keep in mind that this estimate may be very different from reality. Relatively reliable data on the strength of the parent meteoroid material and its distribution would be obtained from physical modelling of the fragmentation during its flight through the atmosphere; \citep{Borovicka2022b, Henych2023}.}, which was $0.0585$~MPa. This discrepancy of an order of magnitude leads us to conclude that this formation process is also highly unlikely.	

Finally, we study the possibility of cluster formation due to thermal stresses. Thermal stresses are mechanical stresses that arise due to thermal expansion in a body with an inhomogeneous temperature field. In a meteoroid moving in interplanetary space, this field arises due to heating by the absorption of solar radiation and cooling by the emission of thermal radiation. The effect of thermal stresses on solar system bodies has been studied by many authors \citep{Kuehrt1984, Tauber1987, Shestakova1997, Tambovtseva1999, Capek2010, Jewitt2010, Delbo2014, Molaro2017, Molaro2020}. 

We first try to estimate whether the thermal stresses are comparable to the expected strength of the parent meteoroid material. For a size of $61$~mm and a circumferential velocity corresponding to the mean ejection velocity, we have a spin frequency of $2$~Hz. Using the analytical model of \citet{Capek2010} with the thermophysical parameters for carbonaceous chondrites, but with a density of $1000$~kg\,m\mt, we obtain an amplitude of the dynamic part of the surface stress tensor of approximately $0.1$~MPa. This is the same value as for the parent meteoroid of the SPE2016 cluster, which is not surprising due to the similar characteristics of the two bodies. Even in the case of the  cluster studied here, the thermal stress is sufficient to exceed the material strength at the surface.

As for the expected magnitude of the ejection velocities, we follow the work of \citet{Molaro2020b}, who estimated the maximum ejection velocities of particles released during the exfoliation of boulders with sizes of a few tens of centimetres on the surface of asteroid (101955)~Bennu as $0.5$~m\,s\mj. This is consistent with the derived ejection velocities for the studied cluster, similarly to the case of the SPE2016 cluster \citep[for a more detailed discussion see][]{Capek2022}.

\subsection{Similarity with the 26 October fireball}
\label{EN261022}

Another fireball with a remarkably similar orbit was recorded by the cameras of the Czech part of the European fireball network (EN) just four days before this cluster was observed. Its orbital elements are also listed in Table~\ref{tab_hel_orbit}. The relatively high value of the error of the large semi-axis is related to the error of the velocity determination. It is also higher than usual due to the large distances of the meteor from all stations.

The Drummond's criterion between both orbits is only 0.078, which is below the threshold value of 0.105 for the meteor shower association \citep{Drummond1981}. Although this is not exactly the case, we used this criterion to illustrate the proximity of the two fireball trajectories. Given the large inclination, their similarity is unusual. Again, there is no association according to the $D_{SH}$ criterion. The radiants differ by a few degrees but almost all orbital parameters are very similar. The only exception is the argument of the perihelion, which differs by almost 16$^{\circ}$. According to the Tisserand parameter, both meteoroids have Halley-type orbits \citep{Tancredi2014}.

\begin{table}
\caption{Physical properties, radiants and orbital elements of the brightest fragment of the cluster and EN261022\_010011 fireball (J2000.0).}             
\label{tab_hel_orbit}
\centering          
\begin{tabular}{l c c} 
\hline
                               &       Fragment no. 1           	   &       EN261022\_010011                \\
\hline
Date                           &       2022-10-30                      &       2022-10-26                      \\
Time [UT]                      &       03:54:04.08                     &       01:00:11.6                      \\
\hline
$H_{BEG}$ [km]                 &       115.9$\pm$0.3                   &       129.44$\pm$0.01                 \\
$H_{END}$ [km]                 &       74.2$\pm$0.2                    &       64.586$\pm$0.009                \\
$l$ [km]                       &       43.8                            &       96.97                           \\
$t_{dur}$ [s]                  &       0.52                            &       1.39                            \\
$M_{MAX}$ [mag]                &       -10.8                           &       -12.34                          \\
$H_{MAX}$ [km]                 &       86.8                            &       73.2                            \\
$m_{phot}$ [kg]                &       1.2e-01                         &       2.73e-01                        \\
$P_{E}$                        &       -5.41                           &       -4.81                           \\
\hline
$\alpha_{G}$ [$^{\circ}$]      &       128.7$\pm$0.3                   &       129.24$\pm$0.03                 \\
$\delta_{G}$ [$^{\circ}$]      &       42.1$\pm$0.2                    &       39.349$\pm$0.008                \\
$V_{G}$ [km/s]                 &       67.1$\pm$1.4                    &       68.14$\pm$0.09                  \\
\hline
a [au]                         &       12$\pm$17                       &       11.6$\pm$1.1                    \\
e                              &       0.92$\pm$0.12                   &       0.914$\pm$0.008                 \\
q [au]                         &       0.961$\pm$0.004                 &       0.99299$\pm$0.00005             \\
$\omega$ [$^{\circ}$]          &       200.0$\pm$1.8                   &       184.27$\pm$0.08                 \\
$\Omega$ [$^{\circ}$]          &       216.3980$\pm$0.0001             &       212.2846$\pm$0.0000             \\
i [$^{\circ}$]                 &       140.6$\pm$0.6                   &       145.47$\pm$0.03                 \\
P [years]                      &       42$\pm$91                       &       39$\pm$5                                \\
$T_{J}$                        &       -0.5$\pm$0.7                    &       -0.55$\pm$0.04                  \\
\hline
$D_{d}$                        &   \multicolumn{2}{c}{0.078}                                           \\
\hline
\end{tabular}
\tablefoot{$H_{BEG}$: Beginning height, $H_{END}$: End height, $l$: Length of trajectory, $t_{dur}$: Duration of meteor shower, $M_{MAX}$: Maximum brightness, $H_{MAX}$: Height of maximum brightness, $m_{phot}$: Photometric mass, $\alpha_{G}$ and $\delta_{G}$: Coordinates of geocentric radiant, $V_{G}$: Geocentric velocity, a: Semimajor axis, e: Eccentricity, q: Perihelion distance, $\omega$: Argument of perihelion, $\Omega$: Longitude of ascending node, i: Inclination, P: Orbital period, $T_{J}$: Tisserand parameter to Jupiter, and $D_{d}$: Drummond's criterion.}
\end{table}

The fireball EN261022\_010011 was detected by four photographic cameras and two video cameras of the EN network. The video cameras that captured its initial part have a similar sensitivity to the AllSky7 cameras. Therefore, it is possible to directly compare the beginning heights of both meteors. We can see that the 26 October fireball started its luminous trajectory at a much higher altitude than the cluster fireball. It was approximately one and a half magnitudes brighter, resulting in more than twice the photometric mass. Due to the consistency with the cluster meteoroid masses, the same luminous efficiency was used \citep{Pecina1983}. According to the $P_{E}$ criterion \citep{Ceplecha1976}, the cluster fireball belongs to class IIIA, while the other bolide belongs to class II. This is consistent with the distribution of the meteors in the cluster, being classes C and B according to the $K_{B}$ criterion. Moreover, as already mentioned, most of the meteors of the cluster fall close to the transition between these two classes. 

Although the orbital relationship of these two events is not proven with certainty, since they were observed close in time to each other and since they both have relatively unusual retrograde orbits with a large inclination to the ecliptic and are not members of any known meteor shower, their orbital similarity is worth noting. The physical properties also do not rule out a relation between the two bodies. It is beyond the scope of this paper to investigate the possible relationship of the two bodies.

\section{Conclusions}
\label{conclusions}

The paper describes a cluster of 22 meteors that was observed over the Baltic Sea region and southern Scandinavia in the morning hours of October 30, 2022. The phenomenon was captured by cameras of the Norwegian/Allsky7 Bolide networks, the brightest meteor was also captured from a large distance by a video camera in the Czech Republic. Despite the large distances of some meteors, the orbits of the individual cluster members could be reconstructed with good accuracy. The observed phenomenon thus ranks among the few well-documented meteor clusters.

Knowledge of the relative positions of individual meteors, their masses and other properties allowed us to study the origin of the meteor cluster. The cluster consisted of a mass-dominant body and numerous smaller fragments.  The distribution of fragments within the cluster corresponds to the release of smaller fragments from the main body at 10.6$\pm$1.7 days before the encounter with the Earth. The ejecta velocities of the fragments ranged from 0.16 to 0.61 ms\mj. 

Three possible scenarios of cluster formation have been studied and disintegration due to thermal stresses appears to be the most likely. The other two processes, i.e. high velocity collision with another body or separation of smaller particles due to rotational fission, are unlikely. 

Interestingly, in a number of properties the studied cluster resembles the September $\epsilon$-Perseid cluster observed in 2016.

\begin{acknowledgements}
This work was supported by the Grant Agency of the Czech Republic grants 20-10907S (data processing and analysis), and the institutional project RVO:67985815 (institutional infrastructure). We thank to reviewer David Clark for his valuable comments that helped us to improve the paper. 
\end{acknowledgements}

\bibliographystyle{aa}
\bibliography{00pkbib}

\end{document}